\documentclass{aa}  
\usepackage{graphicx}
\begin{document}
   \title{Ultraviolet-to-Far Infrared Properties of Lyman Break Galaxies and Luminous Infrared Galaxies at z $\sim 1$}

   \subtitle{}

   \author{
D. Burgarella\inst{1},
Pablo G. P\'erez-Gonz\'alez\inst{2},
Krystal D. Tyler\inst{2},
George H. Rieke\inst{2},
V\'eronique Buat\inst{1},
Tsutomu T. Takeuchi\inst{1,10},
S\'ebastien Lauger\inst{1},
St\'ephane Arnouts\inst{1},
Olivier Ilbert\inst{3},
Tom A. Barlow\inst{4},
Luciana Bianchi\inst{5},
Young-Wook Lee\inst{6},
Barry F. Madore\inst{7,8},
Roger F. Malina\inst{1},
Alex S. Szalay\inst{9}
\and
Sukyoung K. Yi\inst{6}
          }

   \offprints{D. Burgarella}

   \institute{
Observatoire Astronomique Marseille Provence, Laboratoire d'Astrophysique de Marseille,
traverse du siphon, 13376 Marseille cedex 12, France \\
\email{denis.burgarella@; veronique.buat@;tsutomu.takeuchi@; sebastien.lauger@oamp.fr}
         \and
Steward Observatory, University of Arizona, 933 North Cherry Avenue, Tucson, AZ 85721, USA \\
\email{pgperez@; ktyler@; grieke@as.arizona.edu}
         \and
Osservatorio Astronomico di Bologna, via Ranzani, 1 - 47 Bologna - Italy
\email{olivier.ilbert1@bo.astro.it}
         \and
California Institute of Technology, MC 405-47, 1200 E. California Boulevard, Pasadena, CA 91125, USA
\email{tab@srl.caltech.edu}
         \and
Center for Astrophysical Sciences, The Johns Hopkins University, 3400 North Charles St., 
Baltimore, MD 21218, USA
\email{bianchi@skyrv.pha.jhu.edu}
         \and
Center for Space Astrophysics, Yonsei University, Seoul 120-749, Korea
\email{ywlee@;yi@galaxy.yonsei.ac.kr}
         \and
Observatories of the Carnegie Institution of Washington, 813 Santa Barbara Street, Pasadena
91101, USA
\email{barry@ipac.caltech.edu}
         \and
NASA/IPAC Extragalactic Database, California Institute of Technology, Mail Code 100-22, 770 S. Wilson Ave., Pasadena, CA 91125, USA
         \and
Department of Physics and Astronomy, Johns Hopkins University, 3400 North Charles Street,
Baltimore, MD 21218, USA
\email{szalay@tardis.pha.jhu.edu}
         \and
Present address: Astronomical Institute, Tohoku University
Aoba, Aramaki, Aoba-ku, Sendai 980-8578, Japan
\email{szalay@tardis.pha.jhu.edu}
}
   \date{Received ; accepted }

  \abstract
   {}
   {We present the first large, unbiased sample of Lyman Break Galaxies (LBGs) at $z \sim 1$.
Far ultraviolet-dropout (1530 \AA) galaxies in the Chandra Deep Field South have been selected using $GALEX$ 
data. This first large sample in the $z \sim 1$ universe provides us with a high quality reference 
sample of LBGs. }
   {We analyzed the sample from the UV to the IR using $GALEX$, 
$SPITZER$, $ESO$ and $HST$ data.}
   {The morphology (obtained from GOODS 
data) of 75 \% of our LBGs is consistent with a disk. The vast majority of LBGs with an IR detection
are also Luminous Infrared Galaxies (LIRGs). As a class, the galaxies not detected at 24$\mu$m are
an order of magnitude fainter relative to the UV compared with those detected individually, 
suggesting that there may be two types of behavior within the sample. For the IR-bright galaxies, 
there is an apparent upper limit for the UV dust attenuation and this upper limit is 
anti-correlated with the observed UV luminosity. Previous estimates of dust attenuations based on the 
ultraviolet slope are compared to new ones based on the FIR/UV ratio (for LBGs detected at 24 $\mu m$), 
which is usually a more reliable
estimator. Depending on the calibration we use to estimate the total IR luminosity, $\beta$-based 
attenuations $A_{FUV}$ are larger by 0.2 to 0.6 mag. than the ones estimated from FIR/UV ratio. 
Finally, for IR-bright LBGs, median estimated $\beta$-based SFRs are 2-3 times larger than the total SFRs estimated 
as $SFR_{TOT} = SFR_{UV} + SFR_{IR}$ while IR-based SFRs provide values below $SFR_{TOT}$ by 15 - 20 \%.
We use a stacking method to statistically constrain the $24 \mu m$ flux of LBGs non individually detected. The
results suggest that these LBGs do not contain large amounts of dust.}
   {}
   {}

   \keywords{cosmology: observations -- galaxies : starburst -- ultraviolet : galaxies -- infrared : galaxies -- galaxies : extinction
               }
\titlerunning{A Multi-$\lambda$ Analysis of Lyman Break Galaxies at z $\sim$ 1}
\authorrunning{Burgarella D. et al.}

   \maketitle
%

\section{Introduction}

Lyman Break Galaxies (LBGs) are the most numerous objects observed at high redshift ($z > 2 - 3$)
in the rest-frame ultraviolet (UV). The discovery of a large population of LBGs beginning with the 
work of Madau et al. (1996), followed by the spectral confirmation of their redshifts, provided 
the astronomical community with the first large sample of confirmed high redshift galaxies 
(Steidel et al. 1996; Lowenthal et al. 1997). The spectra of bright LBGs (e.g. cB58 by Pettini et 
al. 2000; Teplitz et al. 2000) are remarkably similar to those of local starbursts, indicating that
these objects are forming stars at a high rate. The observed colors of LBGs are redder than expected
for dust-free star-forming objects. This reddening suggests that some dust is present in this
population. However, the amount of dust in LBGs (Baker et al. 2001; Chapman et al. 2000),
and therefore the reddening-corrrected star formation rate ($SFR^c$), is poorly known. Meurer et al. 
(1999), Adelberger \& Steidel (2000) and subsequent papers tried to estimate the amount of dust 
attenuation from the $\beta$ method (Calzetti et al. 1994). However, it has been shown recently 
that this approach only provides rough estimates of the total UV attenuation in local galaxies
(e.g. Buat et al. 2005; Burgarella et al. 2005; Seibert et al. 2005, Goldader et al. 2002 and Bell 2002).

High redshift LBGs are mainly undetected at sub-millimeter (sub-mm) wavelengths, where the 
emission of galaxies is dominated by the dust heated by young stars (Kennicutt 1998). Only the 
most extinguished LBGs are detected by SCUBA (Chapman et al. 2000, Ivison et al. 2005) and we 
have no idea of the dust attenuation for a representative sample. Very recently, Huang et al. 
(2005) observed a population of LBGs at $2 < z < 3$ detected with $SPITZER$. Unfortunately, due 
to the very high redshift, the $SPITZER/MIPS$ observations were not deep enough to detect the 
thermally reradiated emission from very many LBGs at $z \sim 3$ and the $SPITZER/IRAC$ data, 
although deep enough to detect many LBGs, only probe the rest-frame NIR (i.e. no information 
about the dust enshrouded star formation can be inferred). The morphology of LBGs is also a 
matter of debate: early works (e.g. Giavalisco et al. 1996) suggested that LBGs could be ellipsoidal, 
and therefore perhaps the progenitors of ellipticals or of the bulges of spiral galaxies. 
The problem is that we can hardly detect low surface brightness areas at high redshift because 
of the cosmological dimming. For instance, Burgarella et al. (2001) suggest that only compact 
star forming regions could be easily detected in deep $HST$ observations.

On the other hand, observations in the sub-mm range have revealed a population of FIR-bright 
galaxies that might be similar to local (Ultra) Luminous IR galaxies ((U)LIRGs) (Blain et al. 1999)
with $10^{11} L_{\sun} < L_{IR} = L(8-1000\mu m) < 10^{12} L_{\sun}$ for LIRGs 
and $10^{12} L_{\sun} < L_{IR} < 10^{13} L_{\sun}$ for ULIRGs. These objects are likely to
dominate the Cosmic Infrared Background (CIB; Elbaz \& Cesarsky 2003) at high redshift.
The link between LBGs and LIRGs is still an open question: are there two classes of objects
or are they related? If they are two facets of the same population, then we could, 
for instance, correct UV fluxes for the dust attenuation to recover the full Star Formation Density
(e.g. P\'erez-Gonz\'alez et al. 2005). 

The usual way of detecting and identifying LBGs is through the so-called dropout technique,
i.e. the absence of emission in the bluest of a series of bands due to the Lyman break
feature moving redwards with the redshift (e.g. Steidel \& Hamilton 1993, Giavalisco 2002). 
However, it is well known that selection effects can have a very strong influence on the deduced 
characterics of an observed galaxy sample (e.g. Buat et al. 2005; Burgarella et al. 2005). Until now, 
there has been no way to detect a general, unbiased sample of LBGs at low redshift (i.e. $z \leq 2$), 
with the same dropout method very successfully used at $z \geq 2$ because we lacked an
efficient observing facility in the UV range. This was quite unfortunate because the simple fact
that the galaxies are closer to us means that we can access much more information on the
morphology, detect fainter LBGs in the UV and in the IR, and therefore harvest larger samples. $GALEX$
and $SPITZER$ changed this situation and have allowed us to define a large sample of LBGs at $z \sim 1$
in the Chandra Deep Field South (CDFS).  

In this study, we combine the detection  in the UV of true (i.e. with a detected Lyman break) LBGs at
$z \sim 1$ with $GALEX$ and at 24$\mu$m with SPITZER/MIPS. These data let us estimate the total dust
emission and therefore, the dust-to-UV flux ratio, which provides a good tracer of the dust
attenuation in the UV. We also use high spatial resolution images to analyse their morphology.
We are therefore able to perform a complete analysis for the first time on a large LBG sample. 

A cosmology with $H_0 = 70 ~km.s^{-1}.Mpc^{-1}$, $\Omega_M = 0.3$ and $\Omega_{\Lambda}=0.7$ is 
assumed in this paper.
  

\section{The Galaxy Sample}

{\it GALEX} (Martin et al. 2005) observed the CDFS field for 44668 sec (Deep Imaging Survey = DIS)
in both the far ultraviolet (FUV) and the near ultraviolet (NUV). The GALEX field is centered at 
$\alpha$ = 03h32m30.7s, $\delta$ = -27deg52'16.9'' (J2000.0). The {\it GALEX} IR1.1 pipeline
identified 34610 objects within the $1.25^o$-diameter field of view. The GALEX resolution 
(image full width at half Maximum = FWHM) is about 4.5 arcsec in FUV and 6 arcsec in NUV.

This field has also been observed by {\it SPITZER} using MIPS (Rieke et al. 2004) in the 
guaranteed time observing program. The MIPS observations provide about 7 - 8 sources arcmin$^{-2}$ 
at 24 $\mu m$ centered on a $\sim 1.45 \times 0.4 = 0.6 ~deg^2$ field of view. The SPITZER image 
FWHM is about 6 arcsec and almost perfectly matches that of GALEX. 

Redshifts from GOODS (Vanzella et al. 2005) and VVDS (Le F\`evre et al. 2004) are available
at the center of the GALEX field. Part of the $GALEX$ CDFS field was observed by COMBO 17
(Wolf et al. 2004) over $0.5 \times 0.5 ~deg^2$. We made use of COMBO 17 redshifts for objects with
$r < 24.5$. In this range, the quality $\delta z / (1+z)$ remains within $\sigma_z < 0.03$
for 53 \% of the objects. Finally, we obtained photometry from the European Southern Observatory
Imaging Survey (EIS) in U, B, V, R and I. 

We built a sample of LBGs as follows. From the sources with redshifts, we  
selected objects in the rest-frame FUV (i.e. in the observed NUV) that we cross-correlated 
($r = 1$ arcsec) first
with the ground-based optical EIS data, then ($r = 4$ arcsecs) with the MIPS 24 $\mu m$ data. In the resulting
catalogue, we down-selected to objects that COMBO~17 puts in the "GALAXY" class. Then, we 
extracted the sources with redshift $0.9 \leq z \leq 1.3$. Since we wish to study Lyman break galaxies, 
we omitted objects without observed NUV and U flux : we kept only objects down to GALEX NUV 
magnitudes = 24.5 corresponding to the GALEX 80\% completeness level and the U-band limiting
magnitude at U=25.1. We did not use a color-color selection as 
performed by Steidel \& Hamilton (1993). Since we are studying galaxies with known
redshifts, the extra color is not needed to screen out interlopers. 
The selection on the x-axis (i.e. $G-R$ color) 
might bias the sample toward low-reddening LBGs that we wish to avoid 
so we can achieve a more general understanding of the sample. In fact, we find
that the members of our sample fall within the traditional color-color LBG range, or very close
to it, as discussed in Section 3.5.
For galaxies at $z \sim 1$, $GALEX$ FUV corresponds to a rest-frame wavelength of 
$\sim 765$ \AA ~and $GALEX$ NUV corresponds to $\sim 1155$ \AA. The observed FUV and NUV filters 
are therefore in the same rest-frame wavelength ranges as $U$ and $G$ filters used to identify 
Lyman breaks at high redshift 
(e.g. Giavalisco 2002 for a review). The observed FUV-NUV 
color thus gives a clear indication of the Lyman break. We picked the objects with the strongest
indication of a break : 
$FUV-NUV > 2$; the final sample contains 297 LBGs. Of this list, 49 objects
(16.5 \%) have a measured flux above the 80 \% completeness limit of the $SPITZER$ 24 $\mu m$ data
(e.g. P\'erez-Gonz\'alez et al. 2005) and thus at a high enough ratio of signal to noise to
be treated individually. Fig.~1 shows two examples of these LBGs at several 
wavelengths including the two $GALEX$ bands, the B and R EIS bands and the $24 \mu m$ $SPITZER/MIPS$
band. Many additional galaxies were detected by $SPITZER$ but at a weaker level; below we
will describe how we used a stacking technique to probe their properties. 
Our sample of 297 UV-selected LBGs in the redshift range $0.9 \leq z \leq 1.3$ constitutes the 
database that we will study in this work (except for Sect. 3.5). The common $GALEX$ - $SPITZER/MIPS$ - COMBO~17 field of 
view (mainly limited by COMBO~17) is about $0.25~deg^2$, which translates to $\sim 1180$ LBGs
deg$^{-2}$ and $\sim 200$ LBGs deg$^{-2}$ for which we have individual IR detections. 

   \begin{figure*}
   \centering
   \includegraphics[width=15cm]{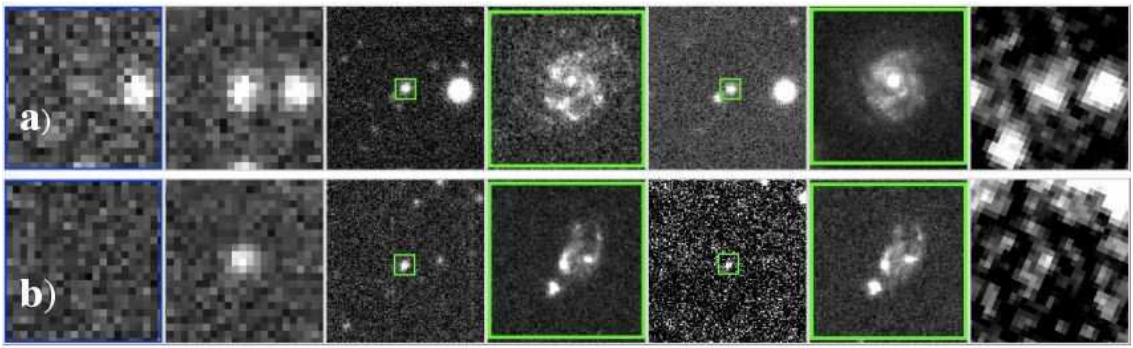}
   \caption{Two galaxies from our LBG sample are shown here, from left to right in $GALEX$ 
FUV and NUV, then $EIS$ B, $HST$ GOODS B, $EIS$ I, $HST$ GOODS I and finally $SPITZER/MIPS$ 
$24 \mu m$ band. For the two galaxies, the leftmost image (blue frame) is below the Lyman 
break at $z \sim 1$ and the galaxy is not visible. The size of GOODS images is 4 $\times$ 
4 arcsec$^2$. The corresponding GOODS field (green frame) is plotted in the large (35 
$\times$ 35 arcsec$^2$) $EIS$ images. a) a LBG classified as a disk-dominated galaxy; 
the more compact object is at $z = 0.546$ from VVDS, it appears reddish and should not 
contribute in ultraviolet and in far infrared; b) a LBG classified as a merger / interacting galaxy.}
              \label{Fig1}%
    \end{figure*}

\section{Lyman Break Galaxies at $z \sim 1$}

Our LBGs provide the first opportunity to study an unbiased sample of LBGs 
systematically at $z \sim 1$. We analyze the UV and IR luminosities of these 
galaxies, measure their morphologies and their 
star formation as revealed by the UV and IR data and finally discuss the implications 
they have for studies centered on higher redshifts.

\subsection{Ultraviolet and Infrared Luminosities}

We find a large range of observed (i.e. un-corrected for dust attenuation) FUV luminosities
($\lambda f_\lambda$ in rest-frame FUV) with $9.3 \leq Log L_{UV} [L_{\sun}] \leq 11.0$. 
The lowest luminosity is set by the limiting magnitude in the U-band at $U=25.1$. 
Below the break, the limiting magnitude amounts to $FUV = 26.0$. Although fainter 
objects are detected in the NUV (down to
$NUV=25.9$), we use a limiting magnitude of $NUV=24.5$ to compute safer $FUV-NUV$ colors 
and therefore perform a safer Lyman Break selection. The average value is $<Log L_{UV} [L_{\sun}]> = 10.2 \pm 0.3 $
for the sample with individual IR detections, and $<Log L_{UV} [L_{\sun}]> = 10.1 \pm 0.3$ for
the rest of the sample.
%
  \begin{figure}
   \centering
   \includegraphics[width=8cm]{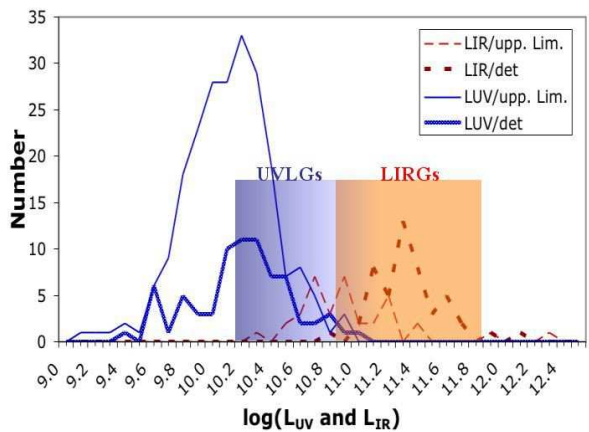}
      \caption{The distribution in observed luminosity presented here corresponds to UV  
luminosities in blue (solid) and IR luminosities in red (dashed). Heavy lines are drawn 
from the population of LBGs with detected counterparts in the $24 \mu m$ MIPS image and 
thin lines to upper limits only. Note that the cut at low $L_{IR}$ is not sharp because 
the $83 \mu Jy$ limit used here is not the detection limit but the 80 \% completeness 
limit. UV luminosities cover the same range for the two samples and reach uncorrected UV 
luminosities of $Log L_{UV} = 11 (in L_{\sun})$. This upper limit is consistent with Adelberger 
\& Steidel range for LBGs at $z \sim 3$ but seems inconsistent with the Balmer-break sample 
at $z \sim 1$. Two areas are shaded. Starting from the left, the first corresponds to the 
range covered by UV Luminous Galaxies (open right ended) and the second one corresponds 
to Luminous IR Galaxies. A few Ultra Luminous IR Galaxies are also detected.}
         \label{Fig2}
   \end{figure}
%

Total IR luminosities ($L_{IR}$) are estimated following the procedure described in P\'erez-Gonz\'alez et 
al. (2005). Briefly, rest-frame 12~$\mu$m fluxes are calculated by comparing the observed mid-IR 
SEDs (including IRAC and MIPS fluxes) of each individual galaxy with models of dust emission (e.g.
Chary \& Elbaz 2001 or Dale et al. 2002). This procedure is meant to cope with the strong 
K-corrections observed in the mid-IR due to the emission from aromatic molecules. We use the formulations of 
Takeuchi et al. (2005) and Chary \& Elbaz (2001) for conversion
from 12$\mu$m flux density to $L_{IR}$. In addition to the intrinsic differences due to the two
calibrations (the former provides $L_{IR}$ lower by 0.2~dex), the conversion from $L_{12 \mu m}$ 
to $L_{IR}$ can introduce errors up to a factor of 2 for individual normal galaxies and 4 for
galaxies with SED variations over the full IRAS sample range 
(Takeuchi et al. 2005; Dale et al. 2005). The effects of these errors
are greatly reduced in this work because we discuss average properties, not those of
individual galaxies; in this case, the uncertainties are likely to be commensurate with
those in the overall conversions to $L_{IR}$ (i.e., still $\sim$ 0.2 dex). There are also 
uncertainties in the luminosity values due to the distance (since we use photometric redshifts with
$\delta z / (z+1) \leq 0.03$), but they are less than 10\% for $L_{UV}$ and $L_{IR}$. 

It is difficult to compare our sample to previously published ones since none was available in this
redshift range before $GALEX$. However, Adelberger \& Steidel (2000) have discussed a Balmer-break
sample at $z \sim 1$. The distributions in $L_{UV}$ and $L_{IR}$ for our sample are shown in Fig.~2.
The lower limit is set by the flux limits but the upper limit of our LBG sample is about the same as
the $z > 3$ one in Adelberger \& Steidel's (2000). However, the upper limit of our sample is
higher by a factor of $\sim 4$ (assuming $H_0 = 70$) than Adelberger \& Steidel's (2000) sample. 

Heckman et al. (2005) defined UV Luminous Galaxies (UVLGs) as galaxies with UV luminosities above
$Log (L_{UV}) = 10.3 ~L_{\sun}$. They found that these UVLGs bear similarities to LBGs, especially a 
sub-sample of compact ones. In our sample, 22.2 \% of the LBGs are UVLGs, 30.6 \% of the LBGs with an IR counterpart are UVLGs. The 83$\mu$Jy
detection limit at 24$\mu$m approximately corresponds to $Log(L_{IR}) \ge 11 L_{\sun}$, that is, 
nearly all the IR-detected LBGs are LIRGs (95.9 \%) and there are 2 ULIRGs (4.1 \%). 
An association between UVLGs and LIRGs was proposed by Burgarella et al. (2005). We confirm here 
this association and extend it to LBGs.

\subsection{The Morphology}

The Great Observatories Origins Deep Survey (GOODS) provides high resolution and high signal-to-noise
images of some of our LBGs, which can be used to study their morphology in the rest-frame B band.
An advantage of our low redshift sample of LBGs is that the images extend to low surface brightness
and hence morphologies can be determined well. We compute the asymmetry and concentration (Fig.~3) 
as in Lauger et al. (2005a) from the objects within the GOODS field which have a signal-to-noise
ratio larger than $S/N \approx 1$ per pixel and whose coordinates are within 2 arcsecs from the GALEX detection.
We were able to obtain the morphology for only 36 LBGS out of our 300 LBGs (about 1/4 of our GALEX + SPITZER
+ COMBO 17 field is covered by GOODS). Fig.~3 
leads us to two conclusions: i) all but one LBG in our sample are located on the disk side of 
the line separating
disk-dominated and bulge-dominated galaxies, and ii) part of them (22 \%) are in the top part of
the diagram, i.e. with an Asymmetry larger than 0.25 and could be interpreted as mergers. 
This kind of quantitative analysis is also applied to higher redshift LBGs, however,
we must be careful in the interpretation because, even if disks are present,
it would be very difficult to detect them due to the cosmological dimming (e.g. Burgarella et al. 
2001). Indeed, in deep spectroscopic observations (e.g. Moorwood et al. 2000, Pettini et al. 2001), the
profiles of the optical nebular lines suggest the presence of disks in some LBGs.

A number of studies have been devoted to galaxy morphology in the redshift range $0.6 \leq z \leq 1.2$. 
At $z \sim 0.7$, most of the works seem to agree that about 60 - 70\% of the 
objects can be classified as disk-dominated galaxies (mainly spirals and Magellanic irregulars) 
and 10 - 20\% as mergers/interacting galaxies. Here spirals are a sub-group of disks which exhibit 
a more symmetric (spiral) structure than irregular-like objects similar to the Magellanic clouds.
Their asymmetry is therefore lower. Lauger et al. (2005b) found about 70 - 80\% of 
disk-dominated galaxies at $z \sim 1$. Zheng et al. (2004) studied the $HST$ 
morphology of a sample of LIRGs and also found that a large majority ($\sim 85$ \%) of 
them are associated with disk-dominated galaxies. This conclusion is reached whether the 
selection is in the rest-frame ultraviolet (Wolf et al. 2005) or in the infrared (Bell et al. 2005). 
However, some dispersion due to the cosmic variance might exist (Conselice, Blackburne \& Papovich
2005).   

Therefore, overall it appears that the majority of the star formation at $0.6 < z < 1$ resides in 
disks and about half of it in spirals. The numbers that we draw for our LBG sample at $z \sim 1$ 
are globally consistent: we find that $\sim 22$\% of the LBGs are likely mergers (e.g. Fig.~1b), 
$\sim 75$\% are disks (e.g. Fig.~1a) and only $\sim 3$\% (i.e. 1 galaxy) is possibly a spheroid.
In the cases where our LBGs can also be classified as LIRGS from their IR luminosity, our measured
morphologies are consistent with the LIRG morphology measurements of Zheng et al. (2004). 

%
  \begin{figure}
   \centering
   \includegraphics[width=8cm]{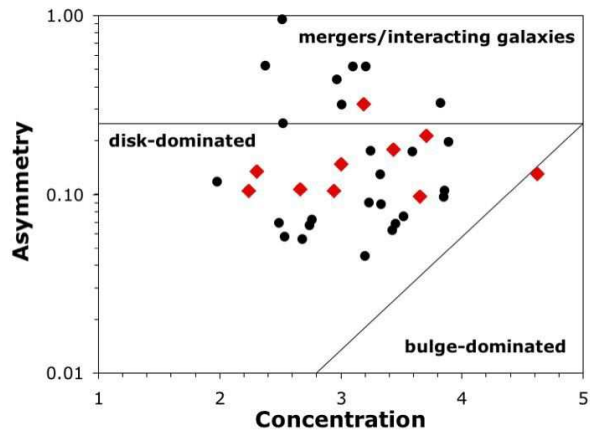}
      \caption{The morphology of the LBG sample is quantitatively estimated from the 
asymmetry and the concentration (sub-sample drawn from a larger CDFS analysis by 
Lauger et al. 2005b). The smaller number of objects presented here is due to the smaller 
field of view of $GOODS$ as compared to ours. Black circless are LBGs without IR 
counterparts while red diamonds represent LBGs with a $24 \mu m$ MIPS detection. The 
line corresponding to the limit between disk-dominated and bulge-dominated galaxies 
(Lauger et al. 2005a). The location in the diagram reflects the morphological type of 
the galaxies: more asymmetrical LBGs (e.g. mergers) are in the top part of the diagram 
($A > 0.25$) while early-type spirals would have $A < 0.1$. The LBG sample is mainly 
dominated (75 \%) by disk-dominated galaxies and the contribution from mergers amounts to $\sim 21$ \%.}
         \label{Fig3}
   \end{figure}
%

\subsection {Ultraviolet Dust Attenuations}

Until now, it has been difficult to estimate the validity of dust attenuation estimates for distant
LBGs, because we had no clear idea of their $L_{IR}$. Adelberger \& Steidel (2000) tried to 
estimate the $800 \mu m$ fluxes of their LBG sample from the $\beta$ method and compared 
the results to observations. However, only the most extreme LBGs can be detected either directly in 
the sub-millimeter range or in the radio range at $1.4 ~GHz$ and therefore could be used in
this comparison. 

In this paper, we use total IR luminosities, $L_{IR}$, and $L_{UV}$ 
to compute the FIR/UV ratio, which is calibrated into FUV dust attenuation $A_{FUV}$ 
(e.g. Burgarella et al. 2005). This method has been shown to provide more accurate dust attenuations 
than those from the UV slope $\beta$. Fig.~4 shows 
an apparent anti-correlation of $A_{FUV}$ with the UV luminosity. 
It is not clear, however, whether this relationship is real or only observational. 
Indeed, in addition to the observational cut at low $L_{FUV}$, the $24 \mu m$ lower 
limiting flux means that we cannot detect individually low-luminosity galaxies with low dust attenuations. 
It is very interesting to note that we do not detect LBGs with both a high UV luminosity 
and a high UV dust attenuation, and this cannot be caused by observational limits. In other words, 
we seem to observe a population of high $L_{UV}$ LBGs (which qualify as UVLGs) with dust 
attenuations similar to UV-selected galaxies in the local universe (e.g. Buat et al. 2005).
UVLG galaxies are LIRGs with the lowest $A_{FUV}$. 

There are now studies of high-redshift LBGs ($z > 2$) with $SPITZER$ (e.g. Labb\'e et al. 2005; 
Huang et al. 2005). However, the results are so far inconsistent: Labb\'e et al. (2005) found that 
LBGs are consistent with low-reddening models while Huang et al. (2005) found more reddened LBGs.
Further observations will resolve these differences and provide a firm basis for comparison with 
our sample.

With the $24 \mu m$ $SPITZER$ flux for the individually detected $z \sim 1$ objects, we can go a step further 
and check how UV dust attenuation estimations carried out from the $\beta$ method 
compare with the better $IR/UV$-based estimates. This comparison has already been 
performed for nearby galaxies (Buat et al. 2005, Burgarella et al. 2005, Seibert et al. 2005 
and references therein). Given the wide use of the $\beta$ method on high redshift LBGs, 
it is useful to compare with our lower-redshift sample. 

Using the equations in Adelberger \& Steidel (2000) (deduced from Meurer et al. (1999)), we
estimate the IR luminosity that is used to compute $A_{FUV}$ and the total luminosity for each 
LBG. We observe a small overestimation of the dust attenuation as compared to the ones estimated 
from $SPITZER/MIPS$ data and the dust-to-UV flux ratio. The $\beta$-based mean dust attenuation
estimated for our $z \sim 1$ LBG sample is $A_{FUV} = 2.76 \pm 0.13$ ($\sigma=1.02$) while the 
dust-to-UV dust attenuation gives $A_{FUV} = 2.16 \pm 0.11$ ($\sigma=0.84$) with $L_{IR}$ from 
Takeuchi et al. (2005) and $A_{FUV} = 2.53 \pm 0.12$ ($\sigma=0.94$) with $L_{IR}$ from Chary \& 
Elbaz (2001). The average value of the two dust-to-UV estimates is consistent with Takeuchi, Buat 
\& Burgarella (2005). The net effect is that total luminosities based on the $\beta$ method are slightly 
larger than the 
actual values and the deduced SFRs are therefore overestimated (see next section). The mean of the
ratios of the $\beta$ vs. FIR/UV $A_{FUV} = 1.31 \pm 0.07$ ($\sigma=0.52$) for Takeuchi et al. (2005)
and $A_{FUV} = 1.09 \pm 0.05$ ($\sigma=0.40$) for Chary \& Elbaz's calibration.

Applying the Kaplan-Meier estimator (and using Chary \& Elbaz's calibrations), we can take upper 
limits into account to estimate mean 
dust attenuations. We find moderate values $<A_{FUV}>=1.36 \pm 0.07$ for $M_{FUV} \leq -22$ 
(5 data points of which 1 is an upper limit) and $<A_{FUV}>=1.08 \pm 0.11$ for $M_{FUV} \leq -21$ i.e.
$L_\star$ at z=3 (35 data points of which 65 \% are upper limits). Using only detections, 
we reached, respectivelly, $<A_{FUV}>=1.54 \pm 0.09$ and $<A_{FUV}>=1.59 \pm 0.18$. Since we only use 
a small number of bright LBGs, this is hardly comparable to the numbers quoted in the previous paragraph. 
However, it suggests that lower $L_{FUV}$ LBGs have higher dust attenuations (see Fig.~4).

%
  \begin{figure}
   \centering
   \includegraphics[width=8cm]{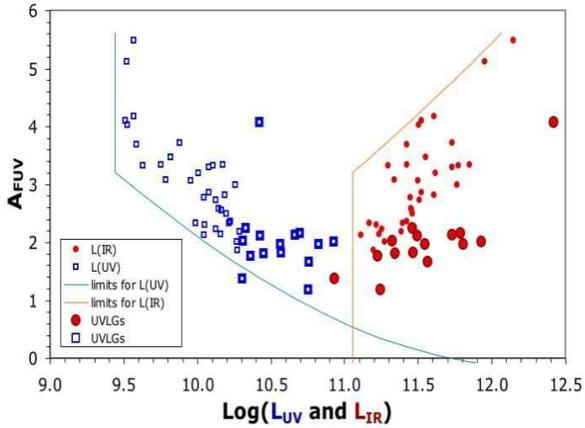}
      \caption{Blue and red symbols are the same objects but luminosities are $L_{UV}$ for 
the former and $L_{IR}$ for the latter. The dust attenuation strongly decreases while $L_{UV}$ 
increases. Part of this apparent correlation might be due to the fact that observational limits 
prevent us from detecting low-luminosity LBGs with low dust attenuation. The clear cut on the upper 
parts of the box cloud cannot be due to observational biases. We do not seem to observe UVLGs 
with high dust attenuations. On the other hand, the more dispersed but well-known increase of 
the dust attenuation with $L_{IR}$ is observed here. Larger symbols correspond to UVLGs 
($Log L_{UV} > 10.3 L_{\sun}$). Most of them have low dust attenuations but one is a ULIRG 
and has $A_{FUV} \sim 4$.}
         \label{Fig4}
   \end{figure}
%

\subsection {Star Formation Rates and Implications for the Cosmic Star Formation Density}

We estimate SFRs for our LBG sample from the IR luminosities and after applying dust corrections 
estimated from $\beta$ and we compare them, in Fig.~5, to the total SFR: $SFR_{TOT} = SFR_{UV} 
+ SFR_{IR}$ where $SFR_{UV}$ is not corrected for dust attenuation. $SFR_{TOT}$ is assumed to be 
the best SFR estimate and we use it as a reference. The first conclusion is that the dispersion is 
much larger for UV $SFR^c$s computed with $\beta$ dust corrections than for IR SFRs. But the median values are also
different: $SFR^c_{UV}$ = 112.3 $\pm$ 33.8 ($\sigma=260.9$) $M_{\sun} .yr^{-1}$ while
$SFR_{IR}$ = 49.9 $\pm$ 13.1 ($\sigma=100.9$) $M_{\sun} .yr^{-1}$ if we use Chary \& Elbaz (2001) and
$SFR_{IR}$ = 30.5 $\pm$ 6.0 ($\sigma=46.6$) $M_{\sun} .yr^{-1}$ if we use Takeuchi, Buat \&
Burgarella (2005). Median $SFR_{TOT}$ for the two above calibrations are, respectivelly, 
$SFR_{TOT}$ = 62.8 $\pm$ 13.2 ($\sigma=102.1$) $M_{\sun} .yr^{-1}$ and 
$SFR_{TOT}$ = 41.1 $\pm$ 6.3 ($\sigma=48.6$) $M_{\sun} .yr^{-1}$. As expected, for LBGs detected in IR,
$SFR_{IR}$ is therefore a better estimate. About 22 \% of our LBGs with 
an IR detection have $SFR_{TOT} > 100$ $M_{\sun} .yr^{-1}$ (using Chary \& Elbaz's calibration) 
as compared to less than 1 \% in Flores et al.'s (1999) galaxy sample which confirms that our LBGs are 
forming stars very actively. However, none of the LBGs undetected at $24 \mu m$ is above 
$SFR_{TOT} > 100$ $M_{\sun} .yr^{-1}$.

The higher SFRs reached when dust attenuations are computed with the UV slope 
$\beta$ (depending on the $L_{IR}$ calibration, +79 to +173\%) lead to an overestimated 
contribution of LBGs to the Cosmic Star Formation Density if the same quantitative 
difference exists at higher redshift. However, Takeuchi, Buat \& Burgarella (2005) 
showed that the current assumption of a constant dust attenuation does not seem 
to be verified. The increase of the mean $A_{FUV}$ from 1.3 to 2.3 from $z=0$ to $z=1$ 
means that, for a given observed FUV luminosity density, the dust-corrected star formation 
density would vary. Note that those mean dust attenuations cannot be compared with the
numbers given for the Kaplan-Meier estimates which are biased toward large $M_{FUV}$ LBGs
while fainter LBGs seem to have larger dust attenuations in our sample. 
Although FIR data are not always available, it is very important 
that one is aware of these uncertainties when using Star Formation Densities derived 
from UV values corrected from the $\beta$ method for LBGs with high dust attenuations, 
especially at high redshift where 
we have a very poor knowledge of actual attenuations. These ambiguities may be 
reduced by further study of Spitzer data, and with Herschel.

%
  \begin{figure}
   \centering
   \includegraphics[width=8cm]{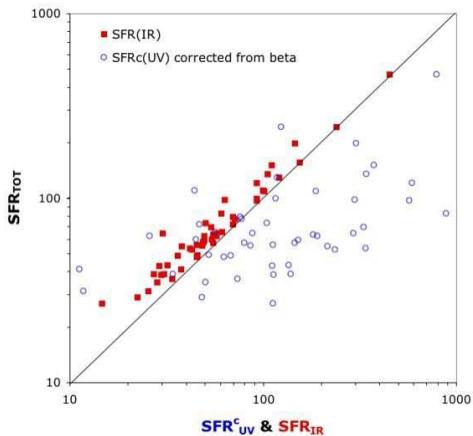}
      \caption{Blue open circles compare $SFR^c_{UV}$ to $SFR_{TOT}$ while red filled boxes compare $SFR_{FIR}$ to $SFR_{TOT}$. Both SFRs are computed from Kennicutt (1998). The median 
$SFR_{FIR}$ is underestimated by about 80 \% and the underestimation increases at lower SFRs, 
which is consistent with the fact that the UV contribution (not accounted for from $L_{IR}$) is 
usually higher at low SFRs than at high SFRs. For this sample of LBG detected at 24 $\mu m$, 
the median value of $SFR^c_{UV}$ is overestimated by 
a factor of 2 - 3 in average with a possible trend for the difference to increase at high SFRs. 
The dispersion of $\beta$-based UV $SFR^c$s ($\sigma_{SFR} \sim 200 M_{\sun} .yr^{-1}$) is much larger than IR-based SFR ($\sigma_{SFR} \sim 30 - 40 M_{\sun} .yr^{-1}$ depending on the calibration into $L_{IR}$).}
         \label{Fig5}
   \end{figure}
%

\subsection{Extension to Galaxies Faint at 24$\mu$m}

So far, most of our arguments have been based on the 49 galaxies individually detected
at 24$\mu$m well above the completeness limit. To put the behavior of these galaxies in 
a broader context, we have determined average infrared flux densities for groups
of galaxies by stacking. Images at 24$\mu$m are shifted to a common center on the basis of
the 24$\mu$m coordinates if the object was well enough detected in that band, or the
coordinates of the optical identification otherwise. We use sigma-clipping to eliminate surrounding sources;
the level at which clipping occurs is adjusted empirically to provide the smoothest
possible sky image. The quoted results are for a level of 5-$\sigma$ for sources detected
at 24$\mu$m and 4-$\sigma$ for undetected ones (see below), but they are not
sensitive to modest adjustments in this level. In general, 
the resulting backgrounds have only weak structure and,
where the sources are fairly bright in the infrared, the 24$\mu$m stacked image is
similar to the point spread function of the instrument. We confirmed that
stacking sources of known flux density gave consistent results. These behaviors validate the
procedure. 

For this study, we used a total of 336 sources, selected similarly to those 
discussed above (but without screening for COMBO-17 type classification). 
We divided the sample into two groups. The first, 
hereafter the undetected group, includes 201 UV 
objects (60\% of the total) for which visual inspection indicated no reliable 24$\mu$m detection. 
The second (40\%), hereafter the detected group, is the objects with evidence 
for an infrared detection; it was in turn 
divided into three equal subgroups according to NUV luminosity. As shown in Table 1, the two groups
have very similar average NUV flux density, 1.89$\mu$Jy for the undetected and
1.63$\mu$Jy for the detected group. The average NUV luminosities are also similar, 
$1.3 \times 10^{44}$ erg/s and $1.1 \times 10^{44}$ ergs/s, respectively. However,
the average infrared flux densities differ by a factor of ten: 13$\mu$Jy for
the undetected group and 143$\mu$Jy for the detected one. There is no significant
difference in average infrared flux density among the subgroups in the detected group.

These results indicate that the LBGs divide into two classes. About 40\% of them
are infrared bright. The average NUV flux density for this group is 1.63$\mu$Jy, so
as measured in $\nu$F$_\nu$, the NUV and 24$\mu$m luminosities are similar. Since there
is a substantial correction to total far infrared luminosity, the infrared component to
the output from young stars is significant, probably accounting for the 
majority of the luminosity for these objects. 
The remaining 60\% are infrared faint: $\nu$F$_\nu$ is about ten times greater in the NUV
than at 24$\mu$m, indicating that their outputs are dominated by the UV. The results of
this paper apply to galaxies like those in the detected group only. 

To explore other possible differences between these classes, we computed the average B 
(i.e. rest-frame NUV) flux densities
for the same two groups and three subgroups. Although it is influenced by other factors,
we take the ratio of B to NUV flux densities (or equivalently the NUV - B color) 
to be an indicator of the level of reddening, and the ratio of 24$\mu$m to B flux density
to measure the relative portion of the luminosity from young stars emerging in
the infrared compared with the UV. The results are in Table 1. First, they demonstrate
that all the galaxies in our selection fall in, or close, to the color-color LBG zone
as adjusted from high z to z $\sim$ 1 (see Giavalisco 2002). There is a trend for LBGs with a high
24/B ratio to present a high B/NUV, which is consistent with the relation for the UV
slope $\beta$ and the FIR/UV ratio found by Meurer et al. (2000) on a sample of local starburst
galaxies. An analysis based on detections is required to check whether Meurer et al.'s law can
be applied safely to those LBGs while we showed in the previous section that it provides dispersed
SFRs for the detected sample. Finally, the amount of dust attenuation
for the undetected group is very low ($A_{FUV} \sim 0.5 - 0.6$ for a mean $Log L_{TOT} \approx 10.5$) 
which corresponds to LBGs with the lowest reddening found by Adelberger and Steidel (2000). This 
very low reddening is consistent with the very blue UV slope $\beta \approx -2.4$. If confirmed, 
this would mean that about half of the LBGs do not contain large amounts of dust. 

%
%

%
\begin{table*}
\caption{Stacking Analysis Results}             
\label{table:1}      
\centering                          
\begin{tabular}{lccccccc}        
\hline\hline                 
 & number & L$_{UV}$  & F(24$\mu$m)  &  F(B)  &  F(NUV)  & F(B)/F(NUV)   & F(24)/F(B)   \\
   &   & 10$^{44}$ergs/sec    & mJy          &  mJy   &   mJy    &               &              \\ 
\hline                        
All detected          &135     &1.11    &0.14       &0.0018  &0.0016   &1.11     &80    \\
low UV               &45     &0.67    &0.16       &0.0014  &0.0012   &1.21     &113   \\
middle UV            &45      &0.99    &0.16       &0.0017  &0.0014   &1.19     &95    \\
high UV              &45      &1.67    &0.14       &0.0023  &0.0024   &0.98     &59    \\
Undetected            &201     &1.33    &0.013      &0.0015  &0.0019   &0.79     &8.7   \\
\hline                                   
\end{tabular}
\end{table*}

\section{Conclusions}

We use multi-wavelength data in the CDFS to define the first large sample of 
Lyman Break Galaxies at $z \sim 1$; $GALEX$ is used to observe the Lyman break. 
Redshifts are taken from spectra and from COMBO 17. 
Quantitative morphologies (Lauger et al. 2005a) are estimated from high 
spatial resolution images. Finally, dust attenuations and total luminosities are computed 
from $SPITZER$ measurements at $24 \mu m$ extrapolated to get the total IR luminosity.

The main results of this analysis are:
   \begin{enumerate}

\item We detect LBGs in the range $9.3 \leq Log L_{FUV} [L_{\sun}] \leq 11.0$, i.e. well into the 
UVLG class as defined by Heckman et al. (2005). For the same objects, $Log L_{IR} \geq 11 L_{\sun}$, 
which means that almost all the LBGs with a $SPITZER$ confirmed detection are LIRGs (and 1 ULIRG)
at $z \sim 1$.

\item LBGs at $z \sim 1$ are mainly disk-dominated galaxies (75 \%) with a small 
contribution of interacting/merging galaxies (22 \%) and a negligible (3 \%) fraction 
are spheroids. The morphologies of our LBG sample are consistent with star-forming 
galaxies.

\item The FUV dust attenuation appears to be anti-correlated with 
the observed FUV luminosity. Part of this correlation might be due to observational 
limits at 24 $\mu m$. However, non-detection of LBG with high $L_{FUV}$ and 
high $A_{FUV}$ cannot be explained by observational biases.

\item About 40\% of our sample of LBGs is detected at 24$\mu$m. These objects also show
evidence for reddening in their UV continua. The remaining 60\% of the sample are
on average an order of magnitude less luminous in the infrared compared with the
rest frame near UV. Their UV continua also appear to be significantly less reddened.

\item Dust corrections of our IR-bright LBG sample computed via the $\beta$ method are over-estimated by 
$\sim 0.2$ mag. if we use Chary \& Elbaz (2001) to compute $L_{IR}$ and by $\sim 0.6$ mag. if we use Takeuchi, 
Buat \& Burgarella (2005).

\item Dust attenuations estimated by the $\beta$ method for such galaxies lead to overestimation of the SFRs at 
$z \sim 1$ by a factor of 2 to 3, depending on the calibration of the $24 \mu m$ flux to $L_{IR}$ while IR-based
SFRs are of the same order as $SFR_{TOT}$.

\item By using the stacking method, we find that LBGs non detected at $24 \mu m$ seem to have very low dust attenuations.

   \end{enumerate}

\begin{acknowledgements}
TTT has been supported by the Japan Society for the Promotion of Sciences 
(April 2004 - December 2005). Support for this work was provided by NASA through Contract Number 
960785 issued by JPL/Caltech. This work is based on observations made with the {\it Spitzer}, 
Space Telescope, which is operated by the Jet Propulsion Laboratory, California Institute of 
Technology under NASA contract 1407. Analysis of the data for this paper was supported by the contract to the MIPS team, 1255094. We also thank the French Programme National Galaxies and the 
Programme National Cosmologie for their financial support. We gratefully acknowledge NASA's support for construction, operation and science analysis for the {\it GALEX} mission developed in cooperation with the Centre National d'Etudes Spatiales of France and the Korean Ministry of Science and Technology. Finally, we thank Emeric Le Floc'h for his help and discussions during this work.
\end{acknowledgements}

\end{document}